**A Praxis of Influence: Framing the Observation and Measurement of Information Power**

Chris Bronk, University of Houston & Rice University

Jason Pittman, University of Maryland Global Campus

Carolyn Semmler, University of Adelaide

**Abstract**

Information power is the capacity to convert data flows into durable shifts in attention, belief, and behavior. We argue that this power has migrated from broadcast persuasion to platform-ized, data-driven operations that fuse computational delivery with cognitive effects. In this context, we define and bound information power within international relations and the information environment while demonstrating why observing and measuring it demands an integrated lens that combines politics (goals and governance), computing (data movement and algorithmic delivery), and psychology (attention, affect, memory, and belief). The article contributes three elements: (1) a triadic analytical framework that specifies the minimum variables and instrumentation needed for study; (2) two crosswalks that map common objectives (persuade, disrupt, shape) and target classes (leaders, elites, publics) to political, computational, and psychological tactics, yielding practical coding heuristics and testable hypotheses; and (3) a McCumber-style cube for information influence that integrates targets, operations, as well as machines (automation and AI) into a single space. The space provides for comparative analysis, data fusion, and effect measurement. Using recent cases across state and commercial platforms, we illustrate how virality, stickiness, and denial of logic exploit fast cognition, why conventional reach metrics understate impact, and where instrumentation should focus. We conclude with a mixed-methods research program coupling computational sensing including large-language-model text mining with experiments and polling. The intention is to move from detecting activity to estimating belief change and decision effects. For policymakers and platforms, the framework yields tractable indicators, audit points, and guardrails that align mitigation with democratic accountability. By making information power observable and measurable, the article offers a shared analytic language and a roadmap for evaluating interventions in a contested information environment.

*Introduction*

More than a decade ago, Joseph Nye offered a revised vision for power relations in which military and economic levers were supplemented by soft power, a term encompassing culture and diplomacy.[1] The idea that international competition is undertaken through the amassing and employment of military and economic power, as well as power over opinion, something in the same vein as soft power, is not new.[2] Advances in computing are changing power relations in all three areas: military, economic and information.[3] Dynamic and rapidly advancing, computing stands as a potential new high ground of global power, much as petroleum or the atom once did. Fears of falling behind in international competition on arms technology or economic growth are increasingly overshadowed by a race for leadership in artificial intelligence (AI). At the same time, human-computer interaction is producing new forms of information influence better understood by inquiry that combines methods and theory from computing as well as politics and psychology.[4]

Computing is transforming human interaction by dramatically shifting how information is created, disseminated and processed, however understanding its application to international relations requires input from multiple scholarly disciplines. The emergence of a global information ecosystem imagined by Vernadsky and Teilhard de Chardin necessitates study of computational information power and how it effects states, elites, and publics. This requires approaches that are: (1) informed of power relations in international relations; (2) able to accurately observe the tableau upon which information power is expended, principally what doctrine calls the information environment, and (3) suffused with approaches for detecting and measuring its employment and impact. This may require dramatically rethinking theoretical guideposts, methods and tools for inquiry, at times selecting unfashionable options, much as Alker and Biersteker did in the mid 1980s, a high-water mark for behavioralist approaches to which they offered a critique.[5]

This is a frameworks paper linking theory and practice, guided by study of information influence in geopolitics studied from political, computational, and psychological disciplinary perspectives. Because it is geopolitical in nature, the International Relations (IR) of information is described and bounded. Next, the space employed for exertion of information power is described as well as the relationship between information, conflict, and cognition. This leads to an explanation of macro variables for understanding the phenomena as well as a heuristic device to guide the process of data collection, analysis, and, finally, correlation. We begin with the idea of power in the international system and how innovation in computation and human-computer interactions is altering information power dynamics in a time of renewed international competition between nation-states.

---

[1] Nye, Joseph S. *The future of power*. Public Affairs, 2011.
[2] Lijphart, Arend. "International relations theory: great debates and lesser debates." *International Social Science Journal* 26, no. 1 (1974).
[3] Onuf, Nicholas. *World of our making: Rules and rule in social theory and international relations*. Routledge, 2012.
[4] Grahn, Hilkka, and James Pamment. "Exploitation of psychological processes in information influence operations: Insights from cognitive science." (2024).
[5] Alker Jr, Hayward R., and Thomas J. Biersteker. "The dialectics of world order: notes for a future archeologist of international savoir faire." *International studies quarterly* 28, no. 2 (1984): 121-142.

*Information and Power in International Relations*

Along with military and economic forms of power, information power is relevant to understanding interactions in the international system.[6] Carr's term, power over opinion, matters a great deal in politics at all levels, especially in democratic states. Foreign influence into domestic politics creates an "intermestic" set of forces on exercise of the democratic franchise.[7] In the employment of information power strategies, computational innovation shapes actors' interactions.[8] For instance, uprisings in Myanmar, Iran, Ukraine, and a host of countries across the Middle East's Arab Spring illustrated the power of user-generated online content platforms (eg. Facebook, Twitter, YouTube) in challenging regime authority.[9] From these events, nondemocratic governments learned an important lesson, that online dissent needed to be met with censorship and suppression.[10] Their leaders implemented technologies to control online expression and restrict connection to popular online platforms or created internal ones for populations seeking similar experiences, albeit with content palatable to those in charge.[11]

From the same set of experiences, leaders in the world's prosperous democracies drew a far different conclusion seeing yet another victory for their newest tools of soft power.[12] Where once these platforms offered an avenue for the US to promote values such as free speech, they have been reversed, and increasingly are used by non-democratic states to subvert norms, increase polarization, and harm societal well-being in democracies. This form of influence, rising to a level of interference, requires a wholesale reappraisal of research programs regarding the attention economy and its ramifications for international relations.[13] Next, however, some bounding of the information space upon which this activity plays out is required.

---

[6] Carr, Edward Hallett. *The twenty years' crisis, 1919-1939: Reissued with a new preface from Michael Cox*. Springer, 2016.
[7] Manning, Bayless. "Congress, the Executive and Intermestic Affairs: Three Proposals." *Foreign Aff.* 55 (1976): 306.
[8] Schmidt, Eric. "AI, great power competition & national security." *Daedalus* 151, no. 2 (2022): 288-298.
[9] Axford, Barrie. "Talk about a revolution: Social media and the MENA uprisings." *Globalizations* 8, no. 5 (2011): 681-686.
[10] Kendall-Taylor, Andrea, Erica Frantz, and Joseph Wright. "The digital dictators: How technology strengthens autocracy." *Foreign Aff*. 99 (2020): 103.
[11] Bronk, Chris. "Collaborating Pariahs: Does the Ukraine War Cement an Adversarial Cyber-Information Bloc?." *Applied Cybersecurity & Internet Governance* 3, no. 1 (2024): 58-77.
[12] Norris, Pippa. "Political mobilization and social networks. The example of the Arab spring." *Electronic democracy* (2012): 53-76.

### *The Tableau of Information Power*

Contemporary information power employs computational and psychological methods to achieve political goals.[14] These are coupled to produce political influence. While often considered regarding democratic elections, influence has manifold forms. Shaping electoral outcomes is but one application. Influence can be designed to alter public views, sow political chaos, or increase polarization.[15] This is manifested in public opinion. For instance, Russia has been able to burnish opinion regarding it in many democracies even as it disrupts those very same countries with elements of its hybrid warfare approach to international competition. A March 2025 US poll in which 34 percent of survey respondents believed Russia to be either an ally or friend,[16] despite the American people underwriting some $128 billion in aid to Ukraine since the country's invasion in 2022.[17] There is a voluminous record of Russian attempts to use a combination of computer hacking, digital propaganda, and elite capture (especially of online personalities) to make it look better or its adversaries and enemies worse.[18]

Where does this struggle of ideology and information take place? Information platforms from Facebook to TikTok have an incredible capacity to draw and hold human attention. Because of this, they have become a significant vehicle for influence undertaken by states and non-state actors. Developed on a funding model of highly specific advertising placements, Meta (owner of Facebook and Instagram) offers a level of targeting to the individual suffused not only by demographic data but also the online behaviors those individuals exhibit upon its platforms.[19] There is an important question for public policy on what constitutes responsible advertising and the public good.[20] The culmination is a propagandist's dream veiled in protections for free speech found in many of the world's most mature democracies. In hindsight it appears that China banned these platforms for good reason, although it has substituted its own versions of them designed to protect the Chinese Communist Party's monopoly on power in the country.[21] It's also necessary to understand how states (and non-state actors as well) develop strategies for using information to influence adversaries. This necessitates visiting the policy and doctrine employed in foreign affairs and international security.

---

[13] Simon, Herbert A (1971). *Designing Organizations for an Information-rich World*. Baltimore, MD: Johns Hopkins University Press. pp. 37–52.
[14] Sun, Ron, ed. *The Cambridge handbook of computational psychology*. Cambridge University Press, 2008.
[15] Weismueller, Jason, Richard L. Gruner, Paul Harrigan, Kristof Coussement, and Shasha Wang. "Information sharing and political polarisation on social media: The role of falsehood and partisanship." *Information Systems Journal* 34, no. 3 (2024): 854-893.
[16] CBS News Poll – February 26-28, 2025. *CBS*. https://d3nkl3psvxxpe9.cloudfront.net/documents/cbsnews_20250302_1.pdf
[17] Masters, Jonathan, and Will Merrow. "Here's How Much Aid the United States Has Sent Ukraine." *Council on Foreign Relations*, March 11, 2025.
[18] Greenberg, Andy. *Sandworm: A new era of cyberwar and the hunt for the Kremlin's most dangerous hackers*. Anchor, 2020.

***Information, conflict and cognition***

States make use of information to achieve their strategic goals through both planned and improvisatory effort. Much about the use of information to influence others in the international system is documented in military doctrine. The lessons here are several. First there is an assumption largely accepted that many states inhabit positions that are neither war nor peace.[22] What was called information warfare and now is labeled operations in the information environment encompass use of information to shape public or elite sentiment and exercise persuasion over thinking at the individual level and in aggregate over populations.[23] The information environment is a defense construct defined by the US military as:

> The aggregate of social, cultural, linguistic, psychological, technical, and physical factors that affect how humans and automated systems derive meaning from, act upon, and are impacted by information, including the individuals, organizations, and systems that collect, process, disseminate, or use information.[24]

---

[19] Bär, D., Pierri, F., De Francisci Morales, G. and Feuerriegel, S., 2024. Systematic discrepancies in the delivery of political ads on Facebook and Instagram. *PNAS nexus*, *3*(7), page247.
[20] Hyman, Michael. "Responsible ads: A workable ideal." *Journal of business ethics* 87 (2009): 199-210.
[21] Kokas, Aynne. *Trafficking data: How China is winning the battle for digital sovereignty*. Oxford University Press, 2023.
[22] Bronk, Christopher. "Between War and Peace: Considering the Statecraft of Cyberspace from the Perspective of the US State Department." In *52nd Convention of the International Studies Association, Montreal, Quebec*. 2011.
[23] Tse, Adam, Swaroop Vattam, Vincent Ercolani, Douglas Stetson, and Mary Ellen Zurko. "A Testbed for Operations in the Information Environment." In *Proceedings of the 17th Cyber Security Experimentation and Test Workshop*, pp. 83-90. 2024.
[24] U.S. Department of Defense. 2023 Department of Defense Strategy for Operations in the Information Environment. November 17, 2023.

These activities cover a broad variety of tactics and methods. They can be physical in nature for instance. Early targets in a variety of US-led military campaigns, telecommunications facilities were once targets for kinetic action, now they may be left in operation to facilitate cyber or information attacks. That Israeli intelligence could make use of Iran's cell phone network to intimidate its military leaders is an example. Information itself may be used in the information environment, with narratives created, shared, and amplified to achieve political goals. Conversely, states and other organizations may attempt to censor or discredit information. Finally, there is a cognitive aspect to information operations that is designed to impact perception, judgement and decision-making.[25] It is within this cognitive segment that political messaging is combined with psychological and computational methods to alter belief among those targeted.

*Cognitive warfare* describes, "the activities conducted in synchronization with other instruments of power, to affect attitudes and behaviors by influencing, protecting, and/or disrupting individual and group cognitions to gain an advantage."[26] It shares a common body of knowledge with psychological operations, but stretches beyond the battlefield to describe interactions with populations beyond combatants engaged in military operations. Cognitive operations embrace taking advantage of cognitive biases, emotions and intellectual shortcuts to achieve political aims, typically in specific segments of a society. Tactics for *cognitive influence* embrace moving targeted individuals to a state of "hot" cognition in which the individual is driven away from deeper or reflective thinking by emotions or information overload.[27] Altering opinion or belief resides at the center of contemporary information power strategy. Detecting the operationalization of that strategy and determining its impact comes into relief through a multifaceted interdisciplinary research program.

With the phenomena considered here, there are of course connectors to larger issues of inquiry. Information power is not exerted in isolation. A military show of force can be considered employment of military power, however, the event becomes part of the information environment as politicians and the press express opinions and concerns about the event. When the US Navy exercised recently with an ultra-long-range air-to-air missile, it demonstrated a capability but also sent an information signal to would be adversaries. Beyond the information environment, the DoD has identified a cyber domain, standing alongside the land, sea, air and space domains as well as an information environment. This information environment is a global information construct unimaginable before the development of the Transmission Control Protocol/Internet Protocol (TCP/IP) stack of technologies. Scholarship regarding cyber and information operations remains largely fixated on one topic or another, when it is likely that they are far more intertwined, something indicated by scholarship from Russia and those on the front lines of contending with Russia.[28]

---

[25] Gomez, Miguel Alberto. "Cyber-enabled information warfare and influence operations: A revolution in technique?." In *Information warfare in the age of cyber conflict*, pp. 132-146. Routledge, 2020.
[26] NATO Allied Command Transformation. "Cognitive Warfare: Strengthening and Defending the Mind." Accessed 12 March 2025.
[27] Thagard, Paul. *Hot thought: Mechanisms and applications of emotional cognition*. MIT press, 2008.

For security contests between actors on the international stage, information assets are used to polarize, destabilize and debase as much as anything else. They are an application of Sun Tzu's maxim, "To subdue the enemy without fighting is the acme of skill." Russia's wars of late are exemplary. While Russia terrifically bungled its attempt to capture Kyiv and failed to prop up its long-faithful ally in Damascus, the country's information operations appear far more successful. This reinforces the belief that traditional warfare is risky, unpredictable, expensive and perhaps a bit old hat. Russia appears far better at undertaking operations other than war, especially by avoiding unrealistic distinctions between cyber and information operations.[29] This is especially true in a hybrid warfare framework that combines information operations, covert action, economic activity and other forms of force. Longtime Russia-watcher Mark Galleotti's thesis on weaponization of all sectors of international activity appears validated through a hybrid lens.[30]

What is meant by this broad weaponization of international interactions? "*[H]ybrid warfare* [author's italics] entails an interplay or fusion of conventional as well as unconventional instruments of power and tools of subversion…to exploit the vulnerabilities of an antagonist and achieve synergistic effects."[31] This form of conflict may include expenditure of effort in information power. It may also include forms of covert action or subversion designed to coerce the targeted state. With the costs of conventional conflict high and often incommensurate with a state's risk tolerance, hybrid activity grows appealing. Where hybrid action intersects with information power objectives is exemplified by low-technology attacks on submarine fiber optic cables, as those cables can sever the large information conduits required for Internet-based communication at scale.[32]

---

[28] Kravchenko, Olena, Vladyslav Veklych, Mykhailo Krykhivskyi, and Tetiana Madryha. "Cybersecurity in the face of information warfare and cyberattacks." *Multidisciplinary Science Journal* 6 (2024).
[29] Akimenko, Valeriy, and Keir Giles. "Russia's Cyber and Information Warfare." *Asia policy* 15, no. 2 (2020): 67-75.
[30] Galeotti, Mark. *The weaponisation of everything: A field guide to the new way of war*. Yale University Press, 2022.
[31] Bilal, Arsalan. "Hybrid Warfare – New Threats, Complexity, and 'Trust' as the Antidote." NATO Review, November 30, 2021.
[32] Kavanagh, Camino. *Wading Murky Waters: Subsea Communications Cables and Responsible State Behaviour*. Geneva: United Nations Institute for Disarmament Research, March 30, 2023.

Information is not just a form of pro-democratic soft power but rather something to be used to bend populations, often unwittingly, to views agreeable to those who wish to push or pull them toward beliefs that they might not have held without some form of influence. What's more, the costs involved in the creation of information decline while sorting and processing it grow. Simon observed this more than five decades ago, suggesting, "In an information-rich world, most of the cost of information is the cost incurred by the recipient."[33] While we might consider this cost to be in time lost to ceaseless "doom scrolling" or watching of micro-videos relatively inconsequential, it may include deeper cognitive effects.[34]

In cybersecurity there is a form of attack that is used to overwhelm the computer servers that deliver web services known as denial of service (DoS). DoS attacks have been around for decades and at times the computer security community appeared to have the problem well in hand, only to be set back by innovation in attacks.[35] They are also inherently measurable; the number of data requests to an Internet host provides a metric for DoS and sets an attack apart from legitimate traffic. What is occurring today are forms of attack that may be characterized as denial of logic or factual truth. Rather than overwhelming servers with a flood of Internet packets, the objective is to overwhelm people with a flood of information designed to produce an emotional reaction. Prior research on Internet memes offered one avenue for doing so, by employing repetition, clever imagery and slogans to produce narrative force able to alter belief. While memes may seem passé, Fox News places one on each of the photos for its top rows of headline stories on its website. For the story "Trump sends a clear message to Supreme Court as federal judges block agenda," that mouthful was given an overlay meme on the accompanying image stating, 'VERY DANGEROUS.'[36] These overlays are a call not only to be informed regarding something but also how to feel about it.

### *An Interdisciplinary Framework for Study of Information Power*

Studying information power in the wake of a computing-based information revolution makes necessary linkages between disciplines for inquiry. Accepting that information power is exerted to achieve political ends, specifically here in the international system, a notional model is offered. It is designed to answer several basic questions. First off, what is the message being delivered by an actor employing the techniques for this sort of power? Second, how is that message conveyed by technology from originator to audience? And finally, how is the message designed to influence the thinking or feelings of its intended target? We have seen attempts to answer these questions from individual disciplines, but the path to more fruitful inquiry likely works its way through at least three of them as well as related ones or even dyadic interdisciplinary communities such as political psychology or computational social science. Three areas appear immediately relevant to understanding information power: politics, psychology, and computing.

---

[33] Simon, Herbert A. "Applying information technology to organization design." *Public administration review* 33, no. 3 (1973): 268-278.

[34] Kosmyna, Nataliya, Eugene Hauptmann, Ye Tong Yuan, Jessica Situ, Xian-Hao Liao, Ashly Vivian Beresnitzky, Iris Braunstein, and Pattie Maes. "Your brain on chatgpt: Accumulation of cognitive debt when using an ai assistant for essay writing task." *arXiv preprint arXiv:2506.08872* (2025).

[35] Schuba, Christoph L., Ivan V. Krsul, Markus G. Kuhn, Eugene H. Spafford, Aurobindo Sundaram, and Diego Zamboni. "Analysis of a denial of service attack on TCP." In *Proceedings. 1997 IEEE Symposium on Security and Privacy (Cat. No. 97CB36097)*, pp. 208-223. IEEE, 1997.

[36] "Trump Demands Supreme Court Step in after Federal Judges Block His Agenda: 'These People Are Lunatics'." *Fox News*, July 15, 2025.

Figure 1: Multidisciplinary approaches to study of Information Power

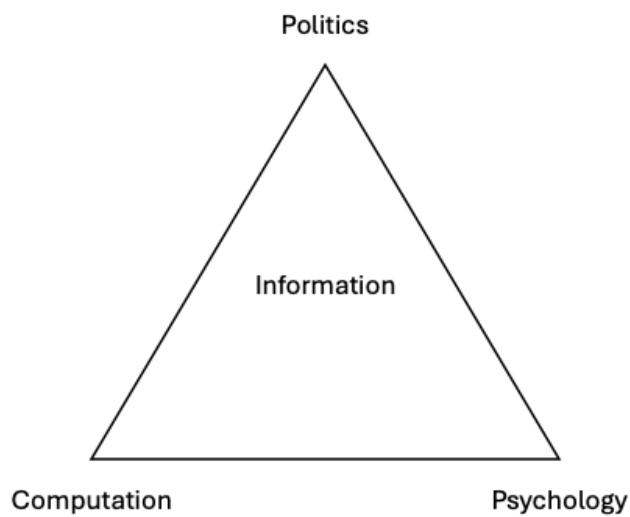

Politics, including political science and international relations, opens an investigational lens on the political behavior of actors from individuals to states. For decades, political scientists have crafted quantitative measurement schemes for studying democracy. Democratic practice, from the voting behavior of citizens to the activity of legislative bodies, has offered the potential to tie inputs in political activity to the outcomes of it. In addition, qualitative methods for study of politics offer explanative power in the role norms, institutions and culture offer in the formation of individual beliefs, actions of leaders, and function of government. From political scholarship, methods and expertise on the study of political activity and goals for information power may be drawn.

Study of computation allows for understanding the movement of data from creator to consumer. Social media platforms are voluminous in the dissemination of data. Consider

Twitter, now X. In 2012, Twitter users, which numbered roughly 100 million, posted 100,000 tweets per minute or 250 million tweets daily. The full flow of data available to researchers then was referred to as a "firehose" by the company.[37] Computational techniques allowed researchers to tame that flood of data. Algorithmic filters can parse enormous quantities of data and detect desired information or even locate unique patterns not considered. They can aid in differentiating human activity online from that of machines. Ideally, computational methods are tools for easing the human burden in synthesizing knowledge from massive seas of data. They are critical for locating signal of desired phenomena from a vastness of unwanted noise.

The third area of scholarship fundamental to understanding information power is psychology. Psychology investigates the causes of human behavior at the individual, group and community level, so although many explanations are sought primarily as individuals interact with the environment and one another, the discipline is also interested in how this happens in groups and broader society. While platforms such as YouTube or Instagram may be assessed for linguistic content or imagery easily enough by machines, that effort does not provide sufficient answer to the matter of how online media makes individuals feel or what they believe. There are episodes captured in media that can shape a nation's sentiment in aggregate because of the cognitive or psychological effect of seeing them. For instance, CBS news anchor Walter Cronkite's grim assessment for victory following the 1968 Tet Offensive fostered doubt among the millions who watched his nightly telecasts. The killing of four students at Kent State University in May 1970 aired repeatedly on television fueled a popular sentiment that the country's commitment to the counterinsurgency in Vietnam was exhausted. From these anecdotes, we can see how images and narratives can shift both individual and collective sentiment. Psychological theory and methods explain how they do.

---

[37] Morstatter, Fred, Jürgen Pfeffer, Huan Liu, and Kathleen Carley. "Is the sample good enough? Comparing data from Twitter's streaming API with Twitter's firehose." In *Proceedings of the international AAAI conference on web and social media*, vol. 7, no. 1, pp. 400-408. 2013.

**Figure 2: Interdisciplinary connective disciplines**

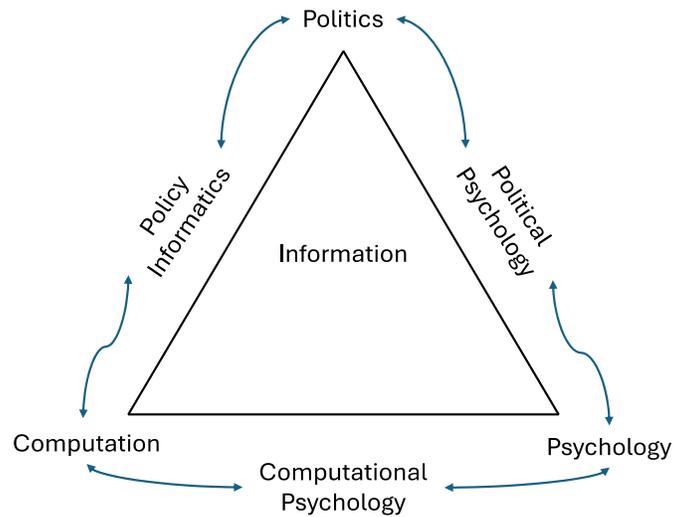

Scholarship regarding social activity has grown increasingly interdisciplinary as single disciplines fail to explain the phenomena of man.[38] While the disciplines can argue over relative importance, the state of inquiry of human-information interactions has reached a point where broad knowledge and narrow alike are desirous. This deeper understanding should also present opportunities to refine new theories and understanding of the processes that govern belief formation within complex information environments - moving theoretical understandings into empirically supported models of influence. Inquiry in information power relations requires deep knowledge across a broad set of areas, necessitating team approaches due to individual limitations of methodological and subject matter expertise.

Politics, Psychology, and Computing are by no means the only disciplines that may be relevant to studying information power relations in the international system. From the humanities, history may have import as well as philosophy.[39] In social science, sociology and anthropology may be useful tools for understanding the employment of narratives at scale. This requirement to study different actors is another key consideration to be addressed.

---

[38] De Chardin, Pierre Teilhard, Julian Huxley, and Bernard Wall. *The phenomenon of man*. London: Collins, 1965.
[39] Abbate, Francesco. "An analysis of informational power transformations: from modern state to the new regime of performativity." *AI & SOCIETY* (2023): 1-12.

***Units of analysis***

Propaganda is generally considered to operate in a one-to-many orientation of messages to receivers.[40] Technology enables new mechanisms for delivery that can tailor individualized messaging based on demographic and social media use data. Mechanical printing presses and aircraft enabled large scale delivery of leaflets on both the battlefield and cities beyond it.[41] After that, radio was identified as a key vehicle for the dissemination of propaganda.[42] Propaganda of the last century or so has depended on a technological component for dissemination. Contemporary propagandists and influencers conduct their narrative campaigns through Internet meme and micro video.[43]

Based in China, Byte Dance's TikTok social micro-video service is considered in US national security circles to be a significant information influence threat.[44] While the issue of political influence delivered by computing technologies is viewed with concern, debate regarding its ramifications to international security come at a time when influencing others is not only a propagandist's objective but also a highly desirable in younger demographic segments of society worldwide. A 2023 survey identified a career as an influencer as most desirable among Americans aged 13-28.[45] With influence a commodity of the attention economy, we want to know whose attention is important to the political influencer operating across international boundaries.[46]

Then there is the matter of leaders and how they are influenced. Much attention has been paid to executive decision-making in democracy, particularly regarding international relations.[47] To what degree online influence activity may shift leadership belief appears a germane question in this vein of inquiry, even if it sounds like the stuff of science fiction. Indeed, philosophers contributed chapters to an edited volume on Christopher Nolan's film *Inception*, which is dedicated to the idea of influence in the subconscious.[48] In cybersecurity, there is an important example on leadership information defense that can be drawn upon. A visit by one of the authors to a large multinational corporation showed how it incorporated a model of leadership protection for its senior officers and their digital connections to both the firm and the outside world. Every attempt to subvert their devices, send nefarious email, or use digital means to target reputation is a corporate cybersecurity problem.[49] How to secure digital communications is now an issue for every major corporate CEO.[50] However, beyond personal cybersecurity, the same program could investigate what information executives choose to consume.

---

[40] Jowett, Garth S., and Victoria O'Donnell. *Propaganda & persuasion*. Sage publications, 2018.
[41] Herz, Martin F. "Some psychological lessons from leaflet propaganda in World War II." *Public Opinion Quarterly* 13, no. 3 (1949): 471-486.
[42] Miller, Clyde R. "Radio and propaganda." *The Annals of the American Academy of Political and Social Science* 213, no. 1 (1941): 69-74.
[43] Farkas, Johan, and Christina Neumayer. "Disguised propaganda from digital to social media." In *Second international handbook of internet research*, pp. 1-17. Springer, Dordrecht, 2018.
[44] Drezner, Daniel W. "How everything became national security: And national security became everything." *Foreign Aff*. 103 (2024): 122.
[45] Morning Consult. *Brands Report: How Brands Can Succeed at Influencer Marketing (Influencer Marketing Trends Report)*, by Amy He, Nicki Zink, Lindsey Roeschke, Claire Tassin, and Kevin Tran, September 2023.
[46] Davenport, Thomas H., and John C. Beck. "The attention economy." *Ubiquity*, no. 1 May 2001.
[47] Reich, Robert B. "'Policy Making in a Democracy," in *The Power of Public Ideas*. Harvard University Press 1988.
[48] Irwin, William. *Inception and philosophy: Because it's never just a dream*. John Wiley & Sons, 2011.

There are also issues of targeted subsets of important figures beyond heads of state. Military officers and enlisted personnel are also potential targets. For instance, before the US-led invasion of Iraq in 2003, Saddam Hussein's generals were persistently engaged in email campaigns designed to subdue them and the forces they commanded.[51] Beyond military personnel, elites are also targeted.[52] Elites hold considerable political, cultural and economic sway in their societies.[53] Many of the new elites are social media influencers or podcasters as well as performing other functions.[54] China has attracted social media influencers with travel to the country in exchange for positive narratives regarding their visits.[55] In the US, influencers and podcasters have accepted substantial financial compensation for their amplification of pro-Russian narratives in the process making themselves the unwitting agents of foreign powers.[56]

At the societal level in democratic states, voting has drawn the attention of information power strategies. While the capacity of social media to aid movements of political protest was first seen in anti-government movements in Myanmar and Iran, with exemplar cases in the Middle Eastern states of the Arab Spring movement.[57] A paired lesson was learned by non-democratic regimes generally and Russia in particular. First, the more that those regimes could seal off their own peoples from outside information, the lower the risk of political destabilization. Second, the major adversarial powers to the community of Western democracies constructed reinvigorated information influence programs often aided by social media firms unwilling to sift malign influence from their voluminous advertising revenues.

---

[49] The author visited a global multinational corporation's cybersecurity operations center in 2022 and saw this function in operation.
[50] Jeffrey Goldberg, "The Trump Administration Accidentally Texted Me Its War Plans," *The Atlantic*, March 24, 2025.
[51] "US Military Send Mixed Messages to Iraqi Generals." *Australian Financial Review*, March 4, 2003.
[52] Higley, John. "Elite theory and elites." In *Handbook of politics: State and society in global perspective*, pp. 161-176. New York, NY: Springer New York, 2010.
[53] Hartmann, Michael. "Elites and power structure." In *Handbook of European societies: social transformations in the 21st century*, pp. 291-323. New York, NY: Springer New York, 2009.
[54] Lasser, Jana, Segun Taofeek Aroyehun, Almog Simchon, Fabio Carrella, David Garcia, and Stephan Lewandowsky. "Social media sharing by political elites: An asymmetric American exceptionalism." *arXiv preprint arXiv:2207.06313*(2022).
[55] "Xinhua Headlines: Chinese Blockbuster 'Ne Zha 2' Brings Boons beyond Theaters." *Xinhua*, February 18, 2025.
[56] Suderman, Alan, and Ali Swenson. "Right-Wing Influencers Were Duped to Work for Covert Russian Operation, US Says." *AP News*, September 5, 2024.
[57] Wolfsfeld, Gadi, Elad Segev, and Tamir Sheafer. "Social media and the Arab Spring: Politics comes first." *The International Journal of Press/Politics* 18, no. 2 (2013): 115-137.

Russia's Internet Research Agency was able to put weight on the scale of the British referendum on exiting the European Union.[58] It later attempted to influence the 2016 US election and was electronically contained to a degree by US Cyber Command in 2018.[59] Such activity continues today, further enabled by the application of AI methods and even through depositing malign data pools for AI algorithms to ingest in their ever-continuing processing of tokenized, textual training data. How publics are targeted, by nation, belief, or other aspects, is a fundamental part of understanding the employment of information power. We are also witnessing an influence of machines unlike anything ever seen before. This may add up to a matrix model of seeing how influence is operationalized across information outlets regarding narratives, the next section of this paper.

***The Narrative Amplification Machine***

Why political narratives take hold, and others don't is a question without simple answer. Understanding 'virality' in social media influencer activity is parts psychology, computation, and perhaps a dash of advertising tradecraft as well as subject matter expertise in the political, economic or social issues in play.[60] To develop a comprehensive understanding of influence, needed is a holistic view of the methods of influence at multiple levels, once again from leaders and elites to publics. Scholarship needs to consider how to measure the influence of information types from Internet memes to long-form podcasts (among others) in shaping the views of individuals and groups.[61] If we consider information operations a symphony of instruments grouped in sections, then it would make sense to understand how the conductor uses those component parts to present a piece of music designed to emotionally move an audience. However, because so much of this information activity is computerized and automated, what we are probably looking for are instructions on how to employ a *narrative amplification machine* rather than an orchestra.

---

[58] Bastos, Marco, and Dan Mercea. "The public accountability of social platforms: Lessons from a study on bots and trolls in the Brexit campaign." *Philosophical Transactions of the Royal Society A: Mathematical, Physical and Engineering Sciences* 376, no. 2128 (2018): 20180003.
[59] Nakashima, Ellen. "US Cyber Command operation disrupted Internet access of Russian troll factory on day of 2018 midterms." *Washington Post* 27 (2019).
[60] Mills, Adam J. "Virality in social media: the SPIN framework." *Journal of public affairs* 12, no. 2 (2012): 162-169.
[61] Chen, Wei, Carlos Castillo, and Laks VS Lakshmanan. *Information and influence propagation in social networks*. Morgan & Claypool Publishers, 2013.

What we want to locate is a virality potential as well as performance to illustrate which campaigns work and which fizzle in reaching a large audience. But as important as the viral reach of a narrative, image or idea is the concept of which ones 'stick.'[62] Democratic elections are filled with both viral and sticky pieces of information. Cleverness and novelty are important, but how those parameters translate to real influence is another matter. In an example from the pre-Internet past, Walter Mondale criticized Ronald Reagan employing a fast-food burger chain television advertisement catchphrase, "Where's the beef?" The line made the rounds across the 1984 electoral news cycle, but Reagan sailed to overwhelming victory in November.[63] It was viral, but not very sticky. More importantly, it lacked narrative influence on the electorate.

Narrative amplification machinery is designed to convince enough actors, and the right actors to alter an outcome.[64] This can have a dramatic difference depending on the underlying system, laws, and norms of each society and state. A fundamental difference between the narrative machine of a democratic state versus a dictatorial, totalitarian one is in the homogeneity of narrative selection, amplification, and power. In a large, heterogenous, democratic nation state, finding and building core issues of interest to constituencies is the challenge. The information winners may be able to shout their way more loudly to achieve their goals. However, if a particular caucus can coordinate their ideation of a narrative as well as its propagation, that group could alter beliefs. Narrative amplification appears to reward those able to move in lock step and punish those who embrace greater diversity of opinion. In the Internet age, where barriers to international connection are few in many democratic states, a multiplicity of narratives may be found. Conversely, in information denying non-democratic regimes, entire narratives may be eliminated by national, packet inspecting firewalls designed to censor, suppressing discourse on issues unpopular among the leader or leaders.[65]

To employ information power across the contemporary information environment requires new frameworks for understanding information channels and the packaging narratives for them.[66] In addition, there is a need to understand the parameters of interest for intended receivers of narratives. Figure 3 displays two narratives in international relations, one regarding the Chinese "nine dash line" of maritime sovereignty for the South China Sea and a second regarding the reciprocal tariffs implemented by the United States. In both cases, the narrative is a contentious one. China's maritime neighbors disagree with what they view as an invalid maritime boundary extension. Conversely, the policy of reciprocal tariffs represents a US-generated narrative, which invokes an image of an international economic system that slights the US economy and the need for policy which ostensibly rebalances trade to be more beneficial to it. Each of these narratives represent a strategic objective in the international system.

---

[62] Suganami, Hidemi. "Narrative explanation and international relations: Back to basics." *Millennium* 37, no. 2 (2008): 327-356.
[63] Gannon, Frank. "Where's the beef?." *EMBO reports* 6, no. 8 (2005): 689-689.
[64] Phillips, Whitney. "The oxygen of amplification." *Data & Society* 22 (2018): 1-128.
[65] Bunn, Matthew. "Reimagining repression: New censorship theory and after." *History and Theory* 54, no. 1 (2015): 25-44.

**Figure 3: Influence tailoring by information platform**

| | Social media | Television and radio | False or astroturf platforms | State-run/owned media | Influencer activity |
|---|---|---|---|---|---|
| Narrative A: Nine dash line | | | | | |
| Narrative B: Reciprocal tariffs | | | | | |

China's nine-dash line assertion has grown across decades, while the American tariff justification narrative is very recent. The nine-dash line possesses narrative endurance, ostensibly because it has been a top point for the Chinese government's portrayal of its sovereign space. Its inclusion in the Hollywood *Barbie* film leaves to wonder if it made the final cut to curry favor with those in China serving as gatekeepers for the distribution of American films in the country. The removal of Taiwan's flag from the *Top Gun's* weathered leather flight jacket is the reverse; information power is exerted in its suppression, once again to play to the economics of film distribution, a century old form of media, but still relevant, nonetheless.[67] American messaging on tariffs is a piece of narratives espousing a need for the restoration of US economic power. Unlike China, the US does not speak on tariffs or much of anything else with one voice. The international message on tariffs is complicated by their employment to support the political arguments of interest groups.[68] Those arguments are rooted in a deep history of protectionism that gets attention in multiple information sectors, from news media to podcasts.

---

[66] Walter, Nathan, Sheila T. Murphy, Lauren B. Frank, and Lourdes Baezconde-Garbanati. "Each medium tells a different story: The effect of message channel on narrative persuasion." *Communication Research Reports* 34, no. 2 (2017): 161-170.

One item where a measure may still be interesting in a narrative or knowledge claim is the accuracy or validity of that claim. We must think of facts and efficacy of argument in using information power. This is another area where psychology of human perception becomes important.[69] The US Republican Party spent $215 million on a television commercial regarding Kamala Harris's position on transgender rights linking it to longstanding narratives on transgender issues in the United States.[70] What we need to know is what those commercials did in changing attitudes, which is not an easy thing to accomplish, and how they did it. This is but a single example. If we want to rate information operation components and their effectiveness, illuminating paths for study of those exposed to such narratives are needed. Data regarding views and speed of conveyance for messages is interesting but so too is opinion regarding a particular piece of information.

We also need to understand audience, targeting and human factors of influence. As we see narratives reshaped for a variety of digital outlets, we can also expect tweaking of messaging to attract attention in different intellectual and emotional demographics.[71] Figure 4 begins to sketch out who are the buyers for narratives. Since the consumers of information are dedicating their scarce attention to a form of media or digital-mediated interaction, they are the buyers of the information power ecosystem.[72] They trade their time and attention for information.

---

[67] Fang, Jun. "The culture of censorship: State intervention and complicit creativity in global film production." *American Sociological Review* 89, no. 3 (2024): 488-517.
[68] Obst, Thomas, Jürgen Matthes, and Samina Sultan. *What if Trump is re-elected? Trade policy implications*. No. 14/2024. IW-Report, 2024.
[69] Pantazi, Myrto, Mikhail Kissine, and Olivier Klein. "The power of the truth bias: False information affects memory and judgment even in the absence of distraction." *Social cognition* 36, no. 2 (2018): 167-198.
[70] Zane McNeill, "Republicans Spent Nearly $215M on TV Ads Attacking Trans Rights This Election," *Truthout*, November 5, 2024.
[71] Green, Melanie C., and Markus Appel. "Narrative transportation: How stories shape how we see ourselves and the world." *Advances in experimental social psychology* 70, no. 1 (2024).
[72] Simon, Herbert A. "Designing organizations for an information-rich world." *International Library of Critical Writings in Economics* 70 (1996): 187-202.

**Figure 4: Influence across targeted constituencies**

|  | Policy influencers | Economic elites | Elected officials | Conspiracy theorists | AI training algorithms |
|---|---|---|---|---|---|
| Narrative A: Nine dash line |  |  |  |  |  |
| Narrative B: Reciprocal tariffs |  |  |  |  |  |

The creators and managers of information campaigns must align narrative, receiver and tailoring of message in hopes of achieving their information influence goals. We know that information operations must have some form of value as advertising persists and is a chief vehicle for the creation of commerce and wealth in technology hubs in Silicon Valley and other places like it. Much of this activity is economic in nature, but when commercial firms of one country advertise to the persons of another, it takes on character of international relations or even international security. For instance, Chinese retailers *Temu* and *Shein* were, until very recently, two of the biggest ad buyers on Meta's platforms, which include Facebook and Instagram, they are also firmly in the middle of the nascent US-China trade war.[73]

That certain narratives are more easily amplified and endure offers some evidence that they are part of the power dynamic in international relations.[74] But we are still limited in our understanding of how technologies and information are altering power relations in the international system. What is encouraging is that our tools for observation and measurement of ideas, images and narratives continue to advance. Artificial intelligence large language models (LLMs) have ingested enormous quantities of textual and other digital information, increasing the capacity of humans to query the information ecosystem. That said, understanding why some narratives take hold while others do not may require additional study via other avenues, for instance polling. We are still very much in the hunt for mechanisms by which we may locate useful variables in the observation of information power relations. What we need is a jumble of characteristics that can be organized to construct useful heuristics for humans and computers alike.

---

[73] Arriana McLymore. "Temu, Shein Slash Digital Ads as Tariffs End Cheap Shipping from China, Data Show." *Reuters*, April 16, 2025. https://www.reuters.com/business/retail-consumer/temu-shein-slash-digital-ads-tariffs-end-cheap-shipping-china-data-show-2025-04-16/.
[74] Miskimmon, Alister, Ben O'Loughlin, and Laura Roselle. *Strategic Narratives: Communication Power and the New World Order*. New York: Routledge, 2013.

***Variables, Vectors and Interactions***

In information influence, a framework must account for a tremendous amount of ambiguity in some areas along with high fidelity measures of data. Like almost anything in the social sciences, a degree of generalization is a necessity especially in these early stages of understanding a form of power relations for which the praxis marches forward, often at what appears breakneck pace. Nonetheless, throwing down stakes on likely orderings or groupings of data must be conducted. Furthermore, that activity most aim for some form of theoretical completeness as well as the precursors for application of computational tools.

For the information influencer, three general categories of actor and action are offered here: targets, operations and machines. Each one of these areas contains a subset of items. For instance, in *operations*, activities may include persuasion, disruption and shaping. In this frame, the influencer may wish to persuade actors regarding the merits of a particular belief. An example of such activity is the influence campaign to alter voter preference and, on a single issue or choice.[75] In the 2016 US elections Russian information operations targeted Black urban voters via social media to convince them that the Democratic Party candidate Hillary Clinton failed to offer the same benefits to them as Barack Obama had.[76] The message in that campaign was simple, the targeted voters received targeted messages telling them that staying home on election day was a valid option. Beyond persuasion, the influencer may want to disrupt their target(s) using political, computational and psychological means.

| Table 1: Information actions with political, computational and psychological heuristics ||||
| --- | --- | --- | --- |
| | **Politics** | **Computation** | **Psychology** |
| **Persuade** | Political campaigns; Issue-based narratives | Targeted ads; Algorithmic content prioritization | Emotional framing; Priming heuristics |

---

[75] The matter of so-called single-issue voters is relevant to this form of targeting. Why try to convince targeted individuals of a broad set of themes when a single one may do the job of deciding the election?
[76] Badawy, Adam, Aseel Addawood, Kristina Lerman, and Emilio Ferrara. "Characterizing the 2016 Russian IRA influence campaign." *Social Network Analysis and Mining* 9, no. 1 (2019): 31. See also: Vićić, Jelena, and Erik Gartzke. "Cyber-enabled influence operations as a 'center of gravity' in cyberconflict: The example of Russian foreign interference in the 2016 US federal election." *Journal of Peace Research* 61, no. 1 (2024): 10-27.

| | | | |
|---|---|---|---|
| **Disrupt** | Destabilizing discourse; Debasing institutions | Denial of Service (DoS) attacks; Manipulative bots | Cognitive overload; Disinformation-induced confusion |
| **Shape** | Agenda-setting; Creation of dichotomies | Social media manipulation; Data-driven nudges | Identity-based appeals; Fear inducement |

Provided here is a vehicle for breaking down the major variables and vectors of information operations. Table 1 describes what political, computational and psychological tactics may be applied to persuade, disrupt or shape belief. The goal of the influencer is to adjust belief and sentiment to produce belief and action that they desire. Political methods are used for framing of issues in a manner that comports with the objective(s) of the influencer. Computational ones are used to deliver a messaging payload, whether through online advertisements or censorship of contrary narratives. Psychological tactics are designed to produce an emotional response even if that response is apathy. To shape the thinking of disparate groups, say supporters of far-right or far-left candidates for election, the influencer may have a similar desired outcome for both, employing a dramatically differentiated political message for each group while employing similar computational and psychological tools and tactics.[77]

| Table 2: Information actions with political, computational and psychological heuristics | | | |
|---|---|---|---|
| | **Politics** | **Computation** | **Psychology** |
| **Leaders** | Executive decision manipulation; Reordering of political objectives | Biographical information; Behavioral data analytics | Leadership specific narratives; Inflammatory information |
| **Elites** | Capture of corporate, government and media figures | Data-driven influencer targeting | Norm change and issue acceptance |
| **Publics** | Mass propaganda; Policy framing | Viral memes; Microtargeting | Altering group beliefs; emotional issue appeals |

---

[77] Badawy, Adam, Kristina Lerman, and Emilio Ferrara. "Who falls for online political manipulation?." In *Companion proceedings of the 2019 world wide web conference*, pp. 162-168. 2019.

Table 2 acknowledges the differences in entities targeted for influence. National leaders are far different than members of the general population who they govern. An information influence strategy may attempt to alter leadership opinion. Clandestine intelligence operations have collected and passed on useful information to diplomats and covert operatives attempting to alter leadership belief or public expression. We have observed this first-hand with the flattery given by some world leaders to others in order to gain traction for policy objectives. In targeting leaders, the challenge is to know what they find important and how their thought process works. This type of influence is deeply personal. Knowledge of the target is of great importance. Knowing what buttons to push for a leader becomes important, especially those with significant executive power. The elites of a nation-state differ from leaders, especially as they are often found in the higher ranks of institutions with considerable power in political matters. Now they run big parts of the US foreign affairs and intelligence enterprise.[78] Finally, publics can be targeted for influence, especially when they are convinced to take desired stances in the democratic franchise, or more dramatically rise up against their own governments or engage in behavior destructive to their societies.

Combining the objective-centric model in Table 1 with the actor-centric model in Table 2, yields a space of objectives, actors, and actions. An intellectual model popular in cybersecurity can serve as a further heuristic device for understanding these phenomena. Figure 5 is a McCumber Cube. It is a feature of cybersecurity research in which John McCumber combined the foundational principles of the discipline (confidentiality, integrity, and availability) with states of information (storage, processing, and transmission) and safeguards put in place to adhere to the principles and information states.[79] Provided here is a McCumber Cube of the labels identified in Tables 1 and 2.

---

[78] Consider how Russia has developed a coterie of media personalities in the US. Some of those personalities, such as US Director of National Intelligence Tulsi Gabbard cozied up to the now-deposed Assad regime in Syria and Department of State undersecretary Darren Beattie, previously delivered messages via the Russia Today media outlet. Regarding Gabbard, see: Fossett, Katelyn. 2025. "He Went to Syria with Tulsi Gabbard. He Has Some Big Concerns." *Politico*, January 9, 2025. Regarding Beattie, see: Pakhnyuk, Lucy. 2025. "Trump Official Who Shut Down Counter-Disinformation Agency Has Kremlin Ties, Telegraph Reports." *The Kyiv Independent*, June 4, 2025.

[79] McCumber, John. *Assessing and managing security risk in IT systems: A structured methodology*. Auerbach Publications, 2004.

**Figure 5: McCumber Cube for illustration of information influence**

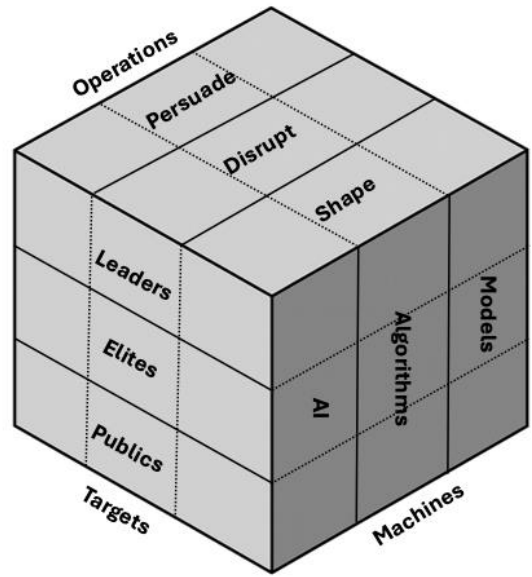

Constructing the exertion of information in a three-dimensional space allows for combining targets, operations, and machines into a single framework. It allows for seeing the interrelation of disparate phenomena. The "machines" plane allows for a deeper computational representation in which computational models and algorithms can be aligned with automation in the form of artificial intelligence methods. This permits for an avenue of inquiry in investigating and categorizing information operations based upon their characteristics and ultimately allows the researcher to mate the framework provided with data from prior or current information operations designed to influence targets of varied type. Many research programs are now successfully collecting data on influence operations undertaken by China, Russia and others (including the United States and its traditional allies in Western Europe and Asia).[80] By applying a unified framework to disparate, but related disciplines may allow data analysis including measurement of effect on targeted populations.

---

[80] Jacobs, Charity S., and Kathleen M. Carley. "# whodefinesdemocracy: Analysis on a 2021 chinese messaging campaign." In *International Conference on Social Computing, Behavioral-Cultural Modeling and Prediction and Behavior Representation in Modeling and Simulation*, pp. 90-100. Cham: Springer International Publishing, 2022. Littell, Joseph, and Nicolas Starck. "Russian influence operations during the invasion of Ukraine." In *International Conference on Cyber Warfare and Security*, pp. 209-XIV. Academic Conferences International Limited, 2023. Alizadeh, Meysam, Jacob N. Shapiro, Cody Buntain, and Joshua A. Tucker. "Content-based features predict social media influence operations." *Science advances* 6, no. 30 (2020): eabb5824.

*Concluding thoughts*

At its point of inception, cognitive science was informed by psychology, computation, and linguistics among other disciplines. With time, we have come to see how information systems are part of international interactions involving cognition. The term "cognitive warfare" has come to encompass the effects of information power campaigns that go beyond the visible manifestations of information to deeper ones involving belief, state of mind, and emotional disposition.[81] For those actors able to employ cognitive tools to achieve their information strategies, the world appears increasingly bent to their will. For scholarship, the challenge is in detecting where information influence is employed and how, particularly in the format of digital information. Herbert Simon offered guidance useful to this process 45 years after he published it. "What we are searching for are relative invariants: regularities that hold over considerable stretches of time and ranges of systems."[82] We want to know what is normal for the digital information ecosystem and what may extend beyond normal.

To understand information power and see its employment will not depend on one approach. Both qualitative and quantitative methods are needed understand it. The influencer may adopt pathways that fuse digital connections as well as real person-to-person ones. Is counting the number of people whose beliefs flip on a particular issue enough to express mastery of observation and measurement of information power? This is likely not the case as influencing leaders and elites may be far more useful to the influencer than publics. One recent case is the rumor of a "kill switch" on Lockheed-Martin F-35 combat aircraft now in the inventories of nine NATO air forces and ordered by several more.[83] In the wake of deteriorating US-European relations in early 2025, numerous news items and social media posts decried a previously unpublished US capability to shut off the F-35. Several weeks after the reports, study of the story shows that a Russian influence operation was largely behind this.[84] This will be the challenge for studying information power, that while news will travel fast the analysis of it will likely take longer. Once again Kahneman's slow thinking comes to the fore in our need to assess the acts we identify as manifestations of information power.

---

[81] Claverie, Bernard, and François Du Cluzel. ""Cognitive warfare": The advent of the concept of "cognitics" in the field of warfare." *Cognitive Warfare: the future of cognitive dominance* (2022): 2-1.
[82] Simon, Herbert A. "Cognitive science: The newest science of the artificial." *Cognitive science* 4, no. 1 (1980): 33-46.
[83] Finnerty, Ryan. 2025. "There Is No Kill Switch: Pentagon Denies F-35 Rumours as Calls Grow to Ditch US Defence Products." *FlightGlobal*, March 18, 2025.
[84] McBride, Julian. 2025. "The F-35 Kill Switch Is the Spare Parts." *The Geopolitics*, March 25, 2025.